\documentstyle[epsfig]{aa}
\begin{document}

\title{A search for radio supernovae and supernova remnants \\
in the region of NGC\,1569's super star clusters}

\author{A.~Greve\inst{1}
\and A.~Tarchi\inst{2,3}
\and S.~H\"uttemeister\inst{4,5}
\and R. de Grijs\inst{6}
\and J.M.~van der Hulst\inst{7}
\and S.T.~Garrington\inst{8}
\and N.~Neininger\inst{2}
}
 
\offprints{A.~Greve}

\institute{Institut de Radio Astronomie Millim\'etrique,
           300 rue de la Piscine, 38406 St.\ Martin d`H\`eres, France
\and Astronomisches Institut der Universit\"at Bonn, Auf dem H\"ugel 71,
D--53121 Bonn, Germany
\and Max-Planck Institut f\"ur Radioastronomie, Auf dem H\"ugel 69,
D--53121 Bonn, Germany
\and Astronomisches Institut der Ruhr-Universit\"at Bochum, 
Universit\"atsstr. 150, D--44780 Bochum, Germany
\and Onsala Space Observatory, S--43920 Onsala, Sweden
\and Institute of Astronomy, University of Cambridge, Madingley Road,
Cambridge CB3 0HA, UK
\and Kapteyn Astronomical Instituut, Postbus 800, 9700 AV Groningen, 
The Netherlands
\and Nuffield Radio Astronomy Laboratories, Jodrell Bank, Macclesfield
Cheshire SK11 9DL, UK
}

\date{received date; accepted date\\
}

\abstract{We have used MERLIN, at 1.4 and 5 GHz, to search for radio
supernovae (RSNe) and supernova remnants (SNRs) in the unobscured
irregular dwarf galaxy NGC\,1569, and in particular in the region of its
super star clusters (SSCs) A and B.  Throughout NGC\,1569 we find some 5
RSNe and SNRs but the SSCs and their immediate surroundings are largely
devoid of non-thermal radio sources.  Even though many massive stars 
in the SSCs are expected to have exploded already, when compared with M\,82
and its many SSCs  the absence of RSNe and SNRs in and near A and B may seem
plausible on statistical arguments.  The absence of RSNe and
SNRs in and near A and B may, however, also be due to a violent and turbulent 
outflow of stellar winds and 
supernova ejected material, which does not provide a quiescent environment 
for the development of SNRs within and near the SSCs. 
}
\titlerunning{A search for SNRs in NGC\,1569}

\maketitle

\section{Introduction} 
For some 150 Myr, the Magellanic-type irregular galaxy NGC\,1569
experienced a starburst of relatively low {\it average} star formation rate
($\sim$\,0.5\,M$_{\sun}$yr$^{-1}$), until approximately 10 Myr ago when the
starburst gradually ceased.  Throughout the body of NGC\,1569 the
starburst has produced many young stars (Greggio et al.  1998, Aloisi et
al.  2001) and a large number of star clusters (Hunter et al.  2000),
together containing a total mass of $\sim 10^{8}$\,M$_{\sun}$.  However,
evidence for {\it recent} and {\it locally} very efficient star formation in
NGC\,1569 comes from the Super Star Clusters (SSCs) {\bf A} and {\bf B}
(Ables 1971, Arp $\&$ Sandage 1985, O'Connell et al.  1994) which are
similar in size and mass to those found in NGC\,1705 (Melnick et al. 
1985), M\,82 (O'Connell et al.  1995, de Grijs et al.  2001), and other
amorphous, irregular, and interacting galaxies.  The SSCs are sites
where in a relatively small volume of $\la$\,50\,kpc$^{3}$ a large
number of stars ($\sim$\,10$^{6}$\,M$_{\sun}$) have spontaneously
formed.  Many massive stars produced in the starburst, either in the body
of NGC\,1569 or in and near the SSCs and intermediate-size clusters,
already ended in supernova (SN) explosions which created bubbles and
kpc-sized loops, an outflow of hot X-ray emitting gas, and a component
of young synchrotron radiation (Waller 1991, Heckman et al.  1995,
Israel $\&$ de Bruyn 1988).  We may therefore expect that NGC\,1569
contains a few radio supernovae (RSNe) and supernova remnants (SNRs), in
particular in the region of the SSCs and the intermediate-size clusters
where higher than average star formation occurred only
$\sim$\,5\,--\,10\,Myr ago (Prada et al 1994, O'Connell et al.  1994,
Origlia et al.  2001) and where some star formation may still go on,
although today the galaxy contains only a few 10$^{6}$\,M$_{\sun}$ of 
locally concentrated molecular gas (Greve et al. 
1996, Taylor et al.  1999).  We have observed NGC\,1569 with MERLIN at
1.4 and 5 GHz in order to search for RSNe and SNRs. 

\subsection{NGC 1569 compared to the prototype starburst galaxy M\,82}

For an investigation of the region around {\bf A} and {\bf B}, a
comparison with the SSC and SNR population and environment in the
prototype starburst galaxy M\,82 is highly relevant, in particular
regarding the issue whether the RSNe and SNRs in M\,82 are associated
with SSCs.  The starburst in M\,82 has produced a large number of SSCs
of which $\sim$\,200 are seen with the HST, both in the active starburst
regions ``M82\,A'' and ``C'' (O'Connell et al.  1995; nomenclature from
O'Connell \& Mangano 1978), and in the more ancient starburst region
``B'' just outside the centre (de Grijs et al.  2001).  These SSCs are
$\ge$\,5\,--\,50 Myr old so that many massive stars have already ended in
a SN explosion.  Although we may expect that some of the approximately
40, resp.  50 RSNe and SNRs in M\,82 detected with MERLIN (Muxlow et al. 
1994) and the VLA (Huang et al.  1994) are associated with SSCs, none,
or at most one, coincides with the SSCs seen with the HST (Golla et al. 
1996).  Similarly, none -- or at most one -- of the $\sim$\,10
H$_\alpha$-bright SNR candidates detected by de Grijs et al.  (2000) in
M82\,B coincides with either bright VLA 8.4 GHz sources, or the optically
bright, slightly evolved SSCs found in large numbers in this region (cf. 
de Grijs et al.  2001). A simple calculation shows that in a 
population of 100 star clusters of ages similar to those estimated for 
M82\,A and containing 10$^{5}$ and 10$^{6}$ 
stars, one would expect to detect between about 5 and $\sim$\,50 type II 
SNRs at any given moment, assuming any reasonable range of initial mass 
functions. The question remains, therefore, why none of the optically-detected
young compact star clusters show any evidence for the presence of SNRs.

The hypothesis brought forward by Golla et al. (1996) for the absence of RSNe
and SNRs in and near SSCs suggests that the visible SSCs of M\,82 are located 
in the foreground and outside appreciable concentrations of interstellar gas 
so that the SN explosions were unable to sweep up gas and form SNRs. They
argue furthermore that there are 1500\,--\,3000 SSCs in M\,82 and that the 
detected RSNe and SNRs are hidden behind dense layers of dust so that the 
associated SSCs are not seen. This argument has apparently gained support from
the recent MERLIN observation (Wills et al.  1998) of H\,{\sc i} absorption 
in the direction of many RSNe and SNRs in M\,82, and from estimates by Mattila
$\&$ Meikle (2001) that the MERLIN-detected sources in M\,82 are hidden behind
dust of $<$A$_{\rm V}$$>$ = 24 ($\sigma$\,$\approx$\,9) mag extinction. 
Evidently, under this condition none of the associated SSCs would be visible. 

Taking M\,82 as example, on statistical arguments we may expect not to
find in NGC\,1569 a short-lived RSN or a SNR in or near the SSCs.  If
indeed the $\sim$\,3$\times$10$^{8}$\,M$_{\sun}$ produced in the
starburst of M\,82 (McLeod et al.  1993) is primarily concentrated in
the predicted 1500\,--\,3000 SSCs, and if the 40\,--\,50 RSNe and SNRs
observed today originated in or near SSCs, then at present {\it at most}
every 1/50\,th to 1/100\,th SSC would be associated with a RSN and
SNR.  Adopting similar conditions for the environment of the SSCs in
NGC\,1569, the chance to observe a RSN or SNR in or near {\bf A} and
{\bf B}, and in and near the intermediate-size clusters, is extremely
small.  This comparison however does not consider the possibility that
SNRs in a dense gas environment, such as in M\,82, may develop
differently than in a Magellanic-type galaxy with generally a small
amount of gas, such as in NGC\,1569.  Evidence and arguments for
different conditions in the interstellar medium in M\,82 have for instance
been advocated by Pedlar et al.  (1999) and Chevalier $\&$
Fransson (2001).  Finally, we may also argue that the environment in and
near SSCs and the intermediate-size clusters may be particularly hostile
at least for the formation of SNRs.  The matter ejected in a SN
explosion in or near the clusters is quickly dispersed because of
stellar winds, nearby SN explosions, and the strong gravi\-tational
field of the clusters.

\begin{figure*}
\psfig{figure=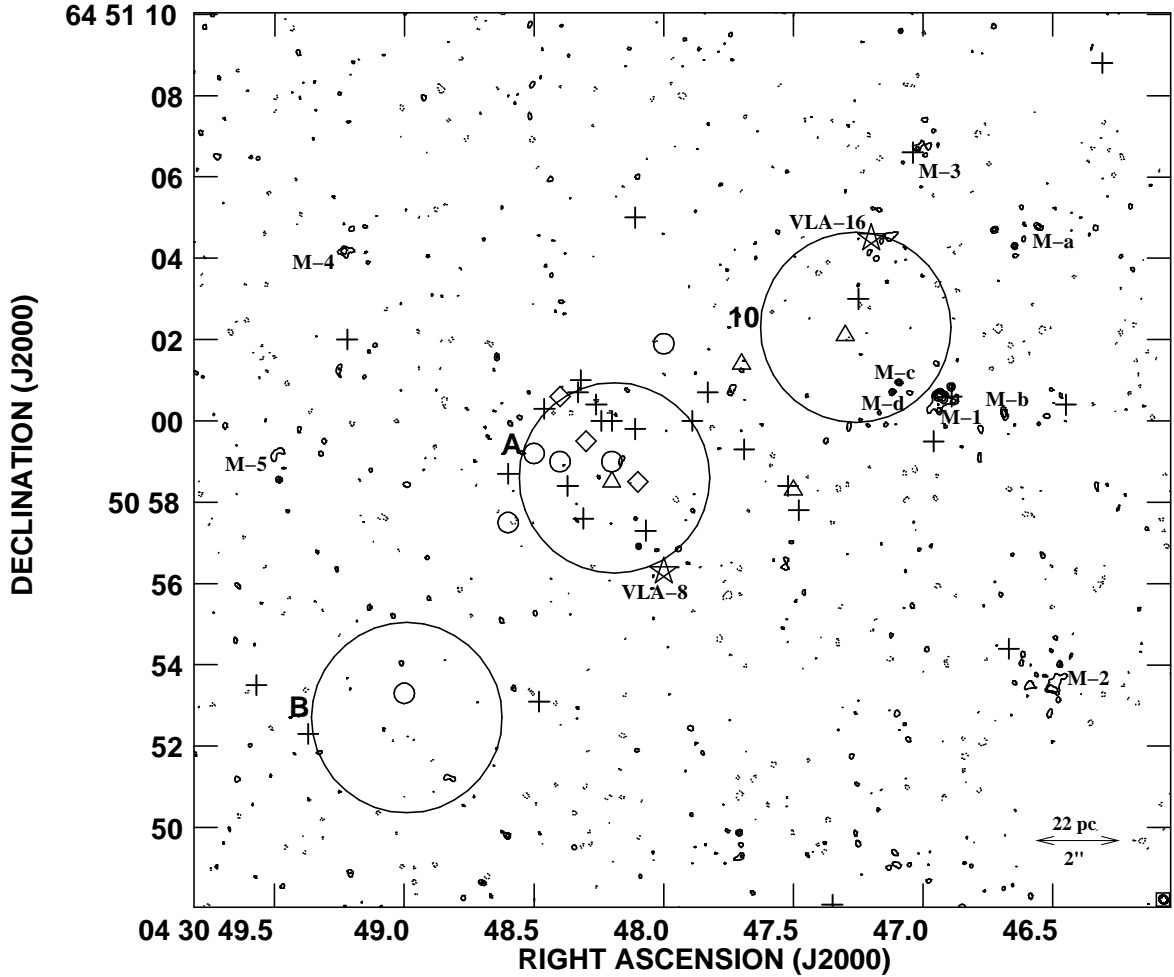,width=16cm,angle=0}
\caption[]{1.4\,GHz MERLIN observation (contours in steps of
30\,$\mu$Jy/beam; first negative and positive contour at 50\,$\mu$Jy/beam) 
of the field around the super star clusters {\bf A}
and {\bf B} and the star cluster {\bf C} (= No\,{\bf 10}) of NGC\,1569
(see Taylor et al.  [1999] for a detailed HST, H$_{\alpha}$ and CO image).  
The encircled areas around the clusters have a diameter of $\sim 50$\,pc. 
The crosses show the positions of star clusters identified by Hunter et
al.  (2000); the open circles, diamonds, and triangles show the
positions of WR objects (S,\,C,\,$\&$ U) observed by Buckalew et al. 
(2000).  \mbox{1$''$ is} equivalent to 11\,pc at the distance of
2.2\,Mpc. \\
\noindent 
The MERLIN-detected sources M--1,\,2,\,3,\,4,\,5 (Table 2) and the tentative
sources M--a,\,b,\,c,\,d (Table 3) are indicated. The stars show the positions
of the VLA-detected sources No\,8 and No\,16 (vdHGI).}
\end{figure*}

Because of the smaller distance to NGC 1569 (2.2\,Mpc; Israel 1988) than
to M\,82 (3.6\,Mpc; cf.  Freedman et al.  1994, Sakai \& Madore 1999),
1.4\,GHz and 5\,GHz MERLIN observations with a resolution of 200\,mas
($\sim$\,2\,pc) and 50\,mas ($\sim$\,0.5\,pc), respectively, are
suitable for a search of RSNe and SNRs. The RSNe and SNRs detected in M\,82 
with MERLIN (Muxlow et al.  1994, Wills et al.  1997) and the VLA (Huang et 
al.  1994) are either unresolved or have diameters of up to $\sim 5$\,pc 
($\sim 400$\,mas); the flux densities measured at 1.4\,GHz and 5\,GHz are
between $\sim 0.5$\,mJy and $\sim 20$\,mJy.  Similar sizes and flux
densities are expected for the RSNe and SNRs in NGC\,1569, if present at
all. 

\section{Observations}
A field of $\sim$\,2$'$$\times$1$'$ centered on the main body of NGC\,1569 was
observed for 20 hours with MERLIN \mbox{(6 antennae)} on Jan 31 and Feb
1, 1999. The observing frequency was 4.994\,GHz 
($\lambda$ = 6\,cm) with a bandwidth of 15\,MHz in both circular 
polarizations. The data were taken in spectral-line mode (32$\times$1\,MHz 
channels).  The QSOs 0552+398 (6.4\,Jy) and 0402+682 (0.18\,Jy) were used as 
flux density and phase calibrators. 

NGC\,1569 was also observed with MERLIN (7 antennae, including the Lovell
telescope) at 1.412\,GHz ($\lambda$ = 21\,cm) for 36 hours in Apr 
(9,\,10,\,11th) and May (3rd) 1999, with a bandwidth of 15\,MHz in both 
polarizations. The passband and relative gain of the antennae were determined 
from observations of QSO 0552+398 (1.75\,Jy); the phases were determined from 
observations of QSO 0402+682 (0.15\,Jy).

Images were produced with the AIPS tasks IMAGR and deconvolved with
CLEAN (Clark 1980).  Table 1 summarizes the details of these
observations.  At 1.4\,GHz, the rms-noise ($\sigma_{\rm n}$) in
source-free fields is consistent with the expected thermal noise level;
for unknown reasons, at 5\,GHz the rms-noise is some 40\,$\%$ too high
as compared to other observations (see Tarchi et al.  2000).  At
1.4\,GHz and 5\,GHz, respectively, sources weaker than $\sim 0.07$
mJy/beam (= 3\,$\sigma_{\rm n}$) and $\sim 0.21$ mJy/beam (=
3\,$\sigma_{\rm n}$) are therefore not detected.

\begin{table}
\caption[]{MERLIN observations of NGC\,1569; values for natural weighting.}
\begin{center}\small
\begin{tabular}{cccc}
\hline
 Frequency  & Beam       & Beam at   & rms--noise ($\sigma$$_{\rm n}$)  \\
            & HPBW       & NGC\,1569$^{\ a)}$   & [$\mu$Jy/beam] \\
\hline
 1.4\,GHz  & 0.21$''$$\times$0.19$''$ & 2.2\,pc$\times$2.0\,pc & $\sim$\,25
       \\
  5\,GHz   & 0.05$''$$\times$0.05$''$ & 0.5\,pc$\times$0.5\,pc & $\sim$\,70
  \\ 
\hline
\multicolumn{4}{@{}l@{}}{%
a) for a distance of 2.2\,Mpc.}
\\
\end{tabular}
\end{center}
\end{table}


\begin{table*}
\caption[]{1.4\,GHz MERLIN-detected sources in NGC\,1569.}
\begin{center}\small
\begin{tabular}{ccccccrc}
\hline
 Source  & RA\,(2000)  & Dec\,(2000) & Peak Flux & Integr. Flux  & Peak
Flux$^{\ a}$ & Spectral & Object$^{\ c}$  \\
         &  [h m s]  & [$\degr$ $'$ $''$] & [mJy/beam] & [mJy] & [mJy] &
Index ($\alpha$)$^{\ b}$ &  \\
\hline
{\bf M--1} & 4 30 46.94 & 64 51 00.6 & 0.16 (0.02) 
(6\,$\sigma$$_{\rm n}$) & 0.57 (0.20) & 1.353$\pm$0.039 & --\,0.02$\pm$0.05 &
Waller No\,2 \\
{\bf M--2} & 4 30 46.51 & 64 50 53.4 & 0.12 (0.02) 
(5\,$\sigma$$_{\rm n}$) & 0.45 (0.25) & 0.848$\pm$0.015 & --\,0.56$\pm$0.05 
& non--thermal \\
{\bf M--3} & 4 30 47.02 & 64 51 06.7 & 0.10 (0.02) (4\,$\sigma$$_{\rm
n}$) & 0.10 (0.04) & 1.107$\pm$0.071 & --\,0.58$\pm$0.10 & non--thermal \\
{\bf M--4} & 4 30 49.20 & 64 51 04.3 & 0.11 (0.02) (4\,$\sigma$$_{\rm n}$) 
& 0.11 (0.04) & 0.411$\pm$0.028 &
0.04$\pm$0.11 & Waller No\,5  \\
{\bf M--5} & 4 30 49.50 & 64 50 59.3 & 0.09 (0.02) (4\,$\sigma$$_{\rm n}$)  &
0.09 (0.04)  & 0.232$\pm$0.016 &
--\,0.02$\pm$0.13 & thermal \\
{\bf M--6} & 4 30 54.13 & 64 50 43.5 & 0.19 (0.02) 
(8\,$\sigma$$_{\rm n}$) & 1.54 (03.0)
& 1.888$\pm$0.027 & --\,0.56$\pm$0.03 & non--thermal \\
\hline
\multicolumn{8}{@{}l@{}}{%
a) VLA observation (van der Hulst et al. 2001), b) S $\propto$
$\nu$$^{{\alpha}}$, 1.4 -- 5 GHz, c) H\,{\sc ii} region number from Waller 
(1991);}
\\
\multicolumn{8}{@{}l@{}}{%
thermal source.}
\\
\end{tabular}
\end{center}
\end{table*}

Since the Wardle telescope was not available for observations, the
shortest baseline at 1.4\,GHz was 7.5\,km (Darnhall--Lovell telescopes)
so that the array was not sensitive to extended structures larger than
$\theta^{\rm 1.4\,GHz}_{\rm max}$ = 5.8$''$ (equivalent to 70\,pc).  At
\mbox{5\,GHz} the shortest baseline was 8.7\,km (Darnhall--Mark2
telescopes) so that, similarly, $\theta^{\rm 5\,GHz}_{\rm max}$ =
1.4$''$ (equivalent to 17\,pc).  However, these fields are large enough
for a search for RSNe, which appear point-like, and SNRs.

We compare the MERLIN observations with 20\,cm and 6\,cm VLA observations made
on Sep 24, 1982 (B-array: 20\,cm $\&$ 6\,cm) and Nov 22, 1983 (A-array:
20\,cm) by van der Hulst et al.  (2001; abbreviated vdHGI). 
The radio images obtained with the VLA were made with 1.18$''$ 
resolution (Gaussian beam) at both frequencies. The spectral indices 
($\alpha$) given below are derived from the VLA measurements.  

At the distance of 2.2\,Mpc, 1$''$ is equivalent to 11\,pc linear scale.
The astrometric accuracy of the MERLIN observation is $\Delta$$\alpha$ =
$\Delta$$\delta$ $\approx$ 0.01$''$, of the VLA observation $\Delta$$\alpha$
= $\Delta$$\delta$ $\approx$ 0.05 to 0.1$''$.

We denote the MERLIN-detected sources by M, the VLA-detected soures (vdHGI)
by VLA.

\begin{figure}
\psfig{figure=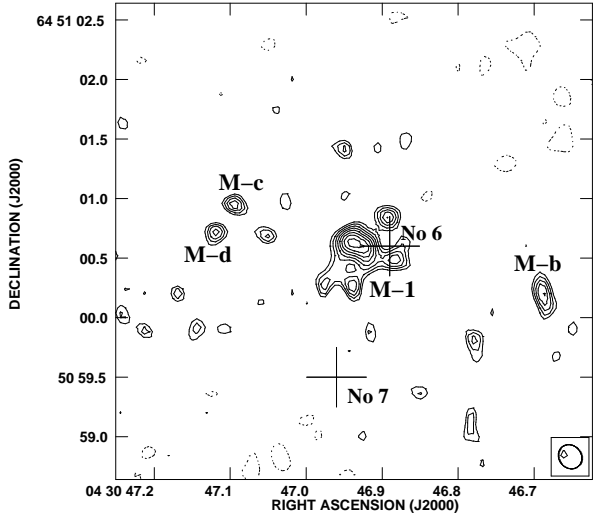,width=8cm,angle=0}
\caption[]{1.4\,GHz MERLIN source M--1 (Table 2). The contour interval 
is 20\,$\mu$Jy/beam with the first negative and positive contour at 
50$\mu$Jy/beam (= 2\,$\sigma$$_{\rm n}$). The synthesized beam is shown in 
the lower right corner. The crosses indicate the center position of the star 
clusters No 6, at $\sim$\,3\,pc to the West, and No 7, at $\sim$\,12\,pc to 
the South (Table 4). The tentative sources M--b,\,c,\,d (Table 3) are 
indicated.} 
\end{figure}

\begin{figure}
\psfig{figure=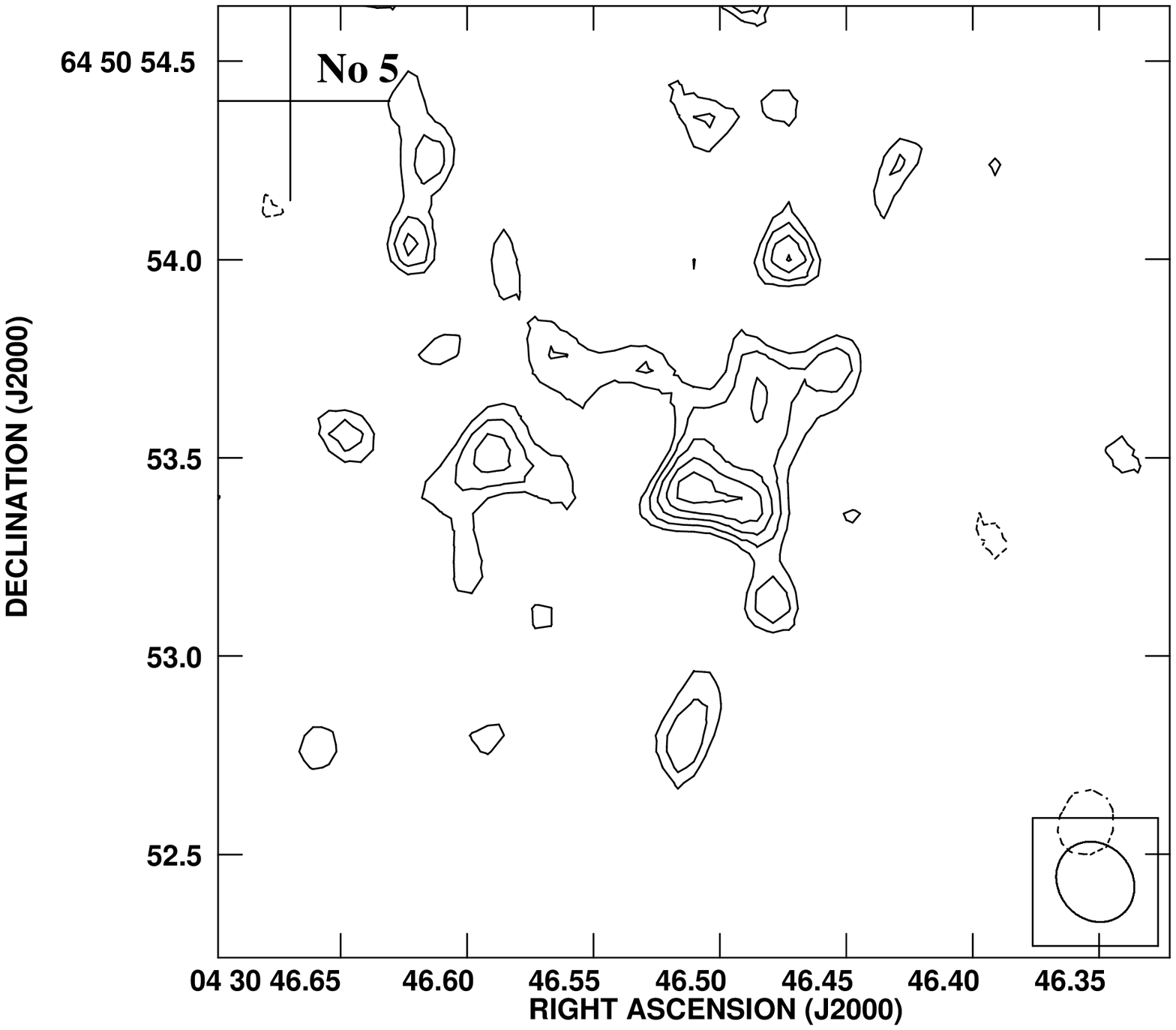,width=8cm,angle=0}
\caption[]{1.4\,GHz MERLIN source M--2 (Table 2). The contour interval 
is 20\,$\mu$Jy/beam with the first negative and positive contour at 
50$\mu$Jy/beam (= 2\,$\sigma$$_{\rm n}$). The synthesized beam is shown in the
lower right corner. The cross indicates the center position of the star 
cluster No 5, $\sim$\,15\,pc to the North-East \mbox{(Table 4)}.} 
\end{figure}

\begin{figure}
\psfig{figure=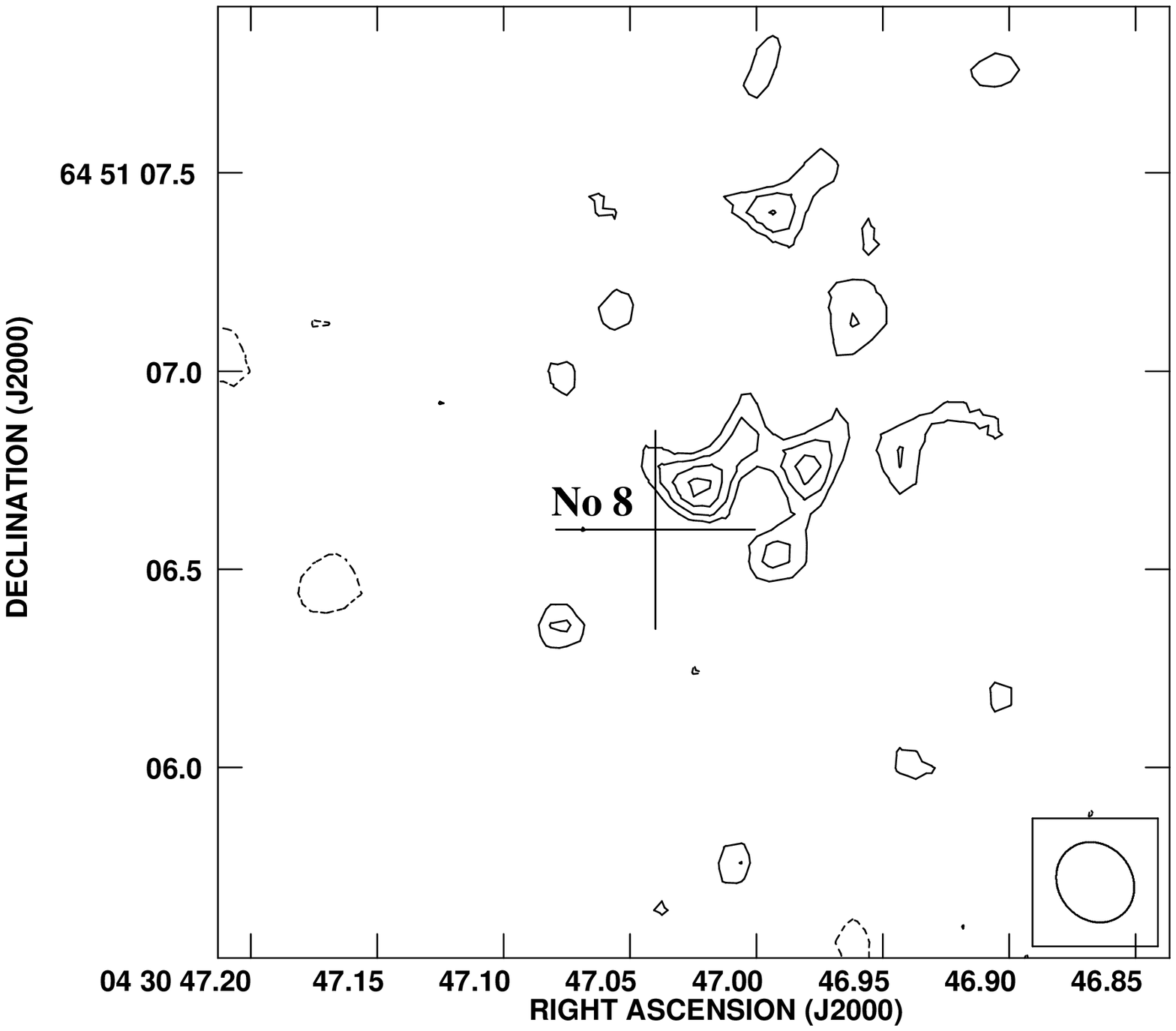,width=8cm,angle=0}
\caption[]{1.4\,GHz MERLIN source M--3 (Table 2). The contour interval 
is 20\,$\mu$Jy/beam with the first negative and positive contour at 
50$\mu$Jy/beam (= 2\,$\sigma$$_{\rm n}$). The synthesized beam is shown in 
the lower right corner. The cross indicates the center position of the star 
cluster No 8, $\sim$\,2\,pc to the South-East \mbox{(Table 4).}} 
\end{figure}

\begin{figure}
\psfig{figure=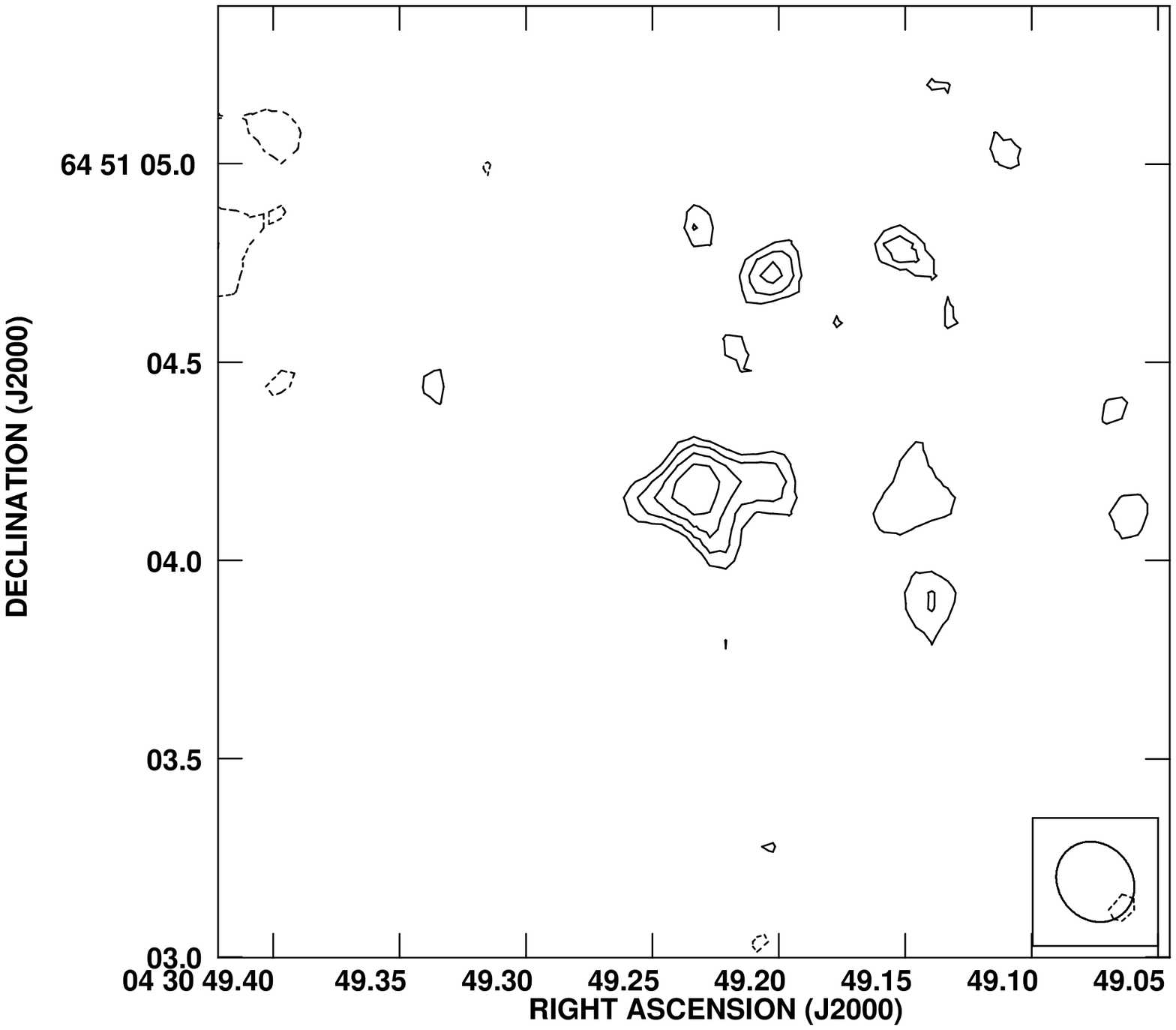,width=8cm,angle=0}
\caption[]{1.4\,GHz MERLIN source  M--4 (Table 2). The contour interval 
is 20\,$\mu$Jy/beam with the first negative and positive contour at 
50$\mu$Jy/beam (= 2\,$\sigma$$_{\rm n}$). The synthesized beam is shown in 
the lower right corner.} 
\end{figure}
\begin{figure}
\psfig{figure=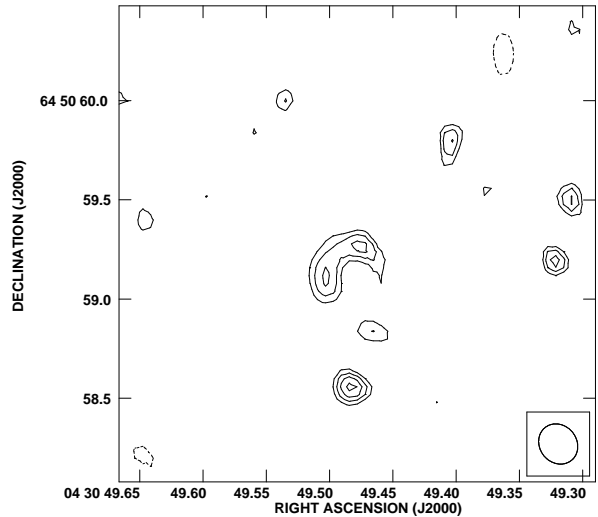,width=8cm,angle=0}
\caption[]{1.4\,GHz MERLIN source M--5 (Table 2). The contour interval 
is 20\,$\mu$Jy/beam with the first negative and positive contour at 
50$\mu$Jy/beam (= 2\,$\sigma$$_{\rm n}$). The synthesized beam is shown 
in the lower right corner.} 
\end{figure}

\section{The detected sources}
Throughout NGC\,1569 we find at 1.4\,GHz six sources, \mbox{M--1} to M--6
listed in Table 2, including their spectral indices $\alpha$ ($S \propto 
\nu^\alpha$, 1.4 -- 5 GHz). The sources are unresolved, except for the 
source M--6.  Because of the (unexplained) high noise of the observations
we do not detect any source above the 3$\sigma$ detection threshold at
5\,GHz. 
The MERLIN sources detected at 1.4 GHz are also present in the VLA 
observations of vdHGI.

Table 3 gives the positions where point-like emission of $\sim$\,0.1\,mJy/beam
peak intensity ($\approx$ 4\,$\sigma$$_{\rm n}$) is observed with MERLIN at 
1.4\,GHz.  Without further observations we are unable to determine whether or 
not this emission is real, and associated with an otherwise identified source.
However, at positions M--b,\,c,\,d the VLA (vdHGI) detects extended emission
at a level of 0.7 mJy/beam extending 2$''$ to the west (possibly containing 
M--b) and 2$''$ to the east (possibly containing M--c,\,d). 

The MERLIN-detected sources are shown in Fig.\,1; detailed images are shown 
in Fig.\,2 to Fig.\,7. The stellar clusters located in the surroundings of the
sources are listed in Table 4 below. In Fig.\,2 to Fig.\,7 the clusters are
identified by the number in the list of Hunter et al. (2000).

We will now discuss the sources relevant to our understanding of the
conditions in the galaxy's interstellar medium in more detail.

\begin{table}
\caption[]{Tentative 1.4\,GHz detections with MERLIN.}
\begin{center}\small
\begin{tabular}{cccc}
\hline
 Source  & RA\,(2000)  & Dec\,(2000) & Figure \\
         &  [h m s]  & [$\degr$ $'$ $''$] & \\
\hline
{\bf M--a} & 4 30 46.64 & 64 51 4.7 & 1 \\
{\bf M--b} & 4 30 46.68 & 64 51 0.1 & 1,2 \\
{\bf M--c} & 4 30 47.08 & 64 51 0.9 & 1,2 \\
{\bf M--d} & 4 30 47.11 & 64 51 0.7 & 1,2 \\
\hline
\end{tabular}
\end{center}
\end{table}

\subsection{Sources in the environment of SSC {\bf A} and {\bf B}}
{\bf VLA--8}\,(Fig.\,1):\ \  this non-thermal radio source with $\alpha 
= -0.24 \pm 
0.10$ is found at $\sim$\,30\,pc to the South-West of SSC--{\bf A} at RA 
04$^{\rm h}$\,30$^{\rm m}$\,48.0$^{\rm s}$, Dec 64$\degr$\,50$'$\,56.3$''$ 
(J2000).  The peak flux densities are S(1.4\,GHz) = $0.479 \pm 0.032$ mJy, 
and S(5\,GHz) = $0.357 \pm 0.022$ mJy. The source is probably extended 
($\sim$\,20\,pc) and of low surface brightness since it is {\it not} seen with 
MERLIN; it may be a SNR based on its spectral appearance and extent (ruling 
out H{\sc ii} regions and RSNe, respectively).  The stellar cluster No\,18
is located at $\sim 10$\,pc to the North of the source (Table 4). In a region
of 2$''$$\times$2$''$ (22\,pc$\times$22\,pc) centered on SSC--{\bf A},
Heckman et al.  (1995) find H$_{\alpha}$ emission of
1\,400\,--\,2\,300\,km\,s$^{-1}$ zero-full-width velocity which may be
evidence of a Balmer-dominated SNR.  In view of the positional
uncertainties, it is unclear whether both observations refer to the same
object. 

\subsection{Sources near cluster {\bf C} and the CO cloud} 
{\bf VLA--16}\,(Fig.\,1):\ \   this source is found $\sim$\,25\,pc to the North
of cluster {\bf C} (No.  10 of Hunter et al. 2000; Table 4).  The source is 
located at RA 4$^{\rm h}$\,30$^{\rm m}$\,47.2$^{\rm s}$, Dec 
64$\degr$\,51$'$\,04.5$''$ (J2000).  It has flux densities of S(1.4\,GHz) = 
$0.681 \pm 0.046$ mJy, and S(5\,GHz) = $0.494 \pm 0.049$ mJy, and a 
non-thermal 
spectral index of $\alpha = -0.27 \pm 0.14$.  This source, {\it not} seen with
MERLIN,  could also be an 
extended ($\sim$\,15\,pc) low surface brightness SNR. 

\noindent
{\bf M--1}\,(Fig.\,2):\ \ when compared with the HST, H$_{\alpha}$, CO and 
star cluster images shown by Taylor et al.  (1999) and Hunter et al.  
(2000), this
source lies within, respectively, 3\,pc and 12\,pc distance from the clusters 
No\,6 and No\,7 (Hunter et al. 2000; Table 4). This source is located at the 
North-Eastern edge of a large molecular cloud complex (No\,3,\,2,\,1 in Taylor
et al.  1999) and coincides with the H\,{\sc ii} region No\,2 of Waller 
(1991).  The source is observed with the VLA (VLA--19) and has a 
thermal spectral index.  M--1 may be double and a second source (cf.  
Fig.\,2) may exist just to the North of the central stellar cluster No\,6 of 
Hunter et al.  (2000).  In this area of the extended H\,{\sc ii} region No\,2
(Waller 1991) lie also the tentative sources M--b,\,c,\,d (Table 3). 

\noindent
{\bf M--2}\,(Fig.\,3):\ \  this source is located at the South-Eastern
side just outside the large molecular cloud complex (3,\,2,\,1) of Taylor et 
al. (1999).  The source lies at a distance of $\sim$\,15\,pc from the cluster 
No\,5 (Hunter et al.  2000; Table 4). The source was detected with the 
VLA (VLA--10) and has a non-thermal spectral index.  It is 
most likely a RSN or a small SNR. 

\noindent
{\bf VLA--11}:\ \   this non-thermal source with $\alpha = -0.74 \pm 0.09$ 
is located some $\sim$\,100\,pc South of {\bf C} and at the South-Western 
side just outside the molecular cloud complex (3,\,2,\,1) 
of Taylor et al.  (1999) at RA 4$^{\rm h}$\,30$^{\rm m}$\,45.79$^{\rm s}$, 
Dec 64$\degr$\,50$'$\,58.3$''$ (J2000), and with flux densities
S(1.4\,GHz) = $0.541 \pm 0.019$ mJy, S(5\,GHz) = $0.223 \pm 0.018$ mJy. 
This source is {\it not} detected with MERLIN. 

\subsection{Other sources} 
{\bf M--3}\,(Fig.\,4):\ \   this source is located $\sim$\,2\,pc to the West 
of cluster No\,8 of Hunter et al. (2000; Table 4).  The source is also 
detected by the VLA (VLA--15). It has a non-thermal spectral index and 
is most likely a RSN or small SNR. 

\noindent
{\bf M--4,\,5}\,(Fig.\,5,\,6):\ \ the source M--4 (VLA--7) is a thermal source
close to the H\,{\sc ii} region No\,5 (Waller 1991); the source M--5 
(VLA--6) is a non--thermal source with some thermal emission.  

\noindent    
{\bf M--6}\,(Fig.\,7):\ \ this source lies in the area of the H\,{\sc
ii} regions 6,\,7\, and 9 identified by Waller (1991), but does not
coincide with any of these objects nor with the H\,{\sc ii} complex
observed by Seaquist et al.  (1976).  The MERLIN 1.4 GHz observation
shows an extended source of $\sim$\,17\,pc diameter.  The source is
detected with the VLA (VLA--1) and has a non-thermal spectrum ($\alpha =
-0.55 \pm 0.02$).  In this area a SNR is detected based on [FeII] line
emission (Labrie $\&$ Pritchet 1998), however, the astrometric precision
of this observation is not sufficient to confidently establish full
correspondence. 

Approximately 500\,pc to the East of {\bf A} and {\bf B}, the VLA observations
reveal 3 to 4 non-thermal sources (VLA--2,\,3,\,4,\,5; S(1.4GHz) $\la$ 
0.4 mJy) in the H\,{\sc ii} region complex at RA $\approx$ 
4$^{\rm h}$\,30$^{\rm m}$\,52$^{\rm s}$, Dec 64$\degr$\,50$'$\,45$''$ (J2000)
(see Seaquist et al. 1976) which are {\it not} seen by MERLIN. If extended, 
these sources are certainly below the detection limit of MERLIN.


\begin{figure}
\psfig{figure=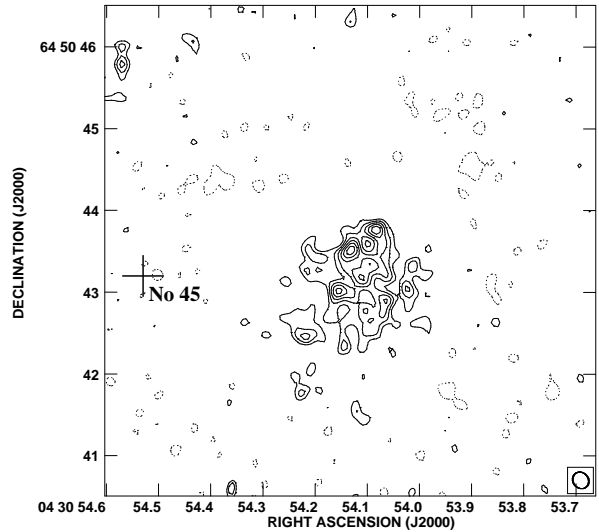,width=8cm,angle=0} 
\caption[]{1.4\,GHz MERLIN observation of source M--6 (Table 2)
associated with the the H\,{\sc ii} regions 7 and 6 identified by Waller
(1991).  The contour interval is in steps of 30\,$\mu$Jy/beam with the
first negative and positive contour at 50\,$\mu$Jy/beam \mbox{(=
2\,$\sigma$$_{\rm n}$).}  The synthesized beam is shown in the lower
right corner.  The cross indicates an intermediate-sized stellar cluster
No\,45 (Table 4).}
\end{figure}

\section{Discussion}

Throughout NGC\,1569 we find 4 to 5 RSNe and/or SNRs, located within an area 
of $\sim$\,300\,pc diameter around the clusters {\bf A, B, C} where active 
star formation occurred until recently, and might still be going on. 
We discuss separately the region near cluster {\bf C} and the associated 
molecular cloud, and the SSCs {\bf A}, {\bf B} and the intermediate-size
clusters and their surroundings.

\subsection{Non-thermal sources near star cluster C and the associated CO 
cloud}

The cluster {\bf C}\,=\,No\,10 is associated with a bright H\,{\sc ii}
region (No\,2, Waller 1991) located at the edge of a CO cloud complex of
$\sim$\,180\,pc extension and 1.4$\times$10$^{6}$\,M$_{\sun}$ total mass
(Taylor et al.  1999, their Figs.\,8\,$\&$\,9).  The sources 
\mbox{M--1,2\,(b,\,c,\,d)}, VLA--10 and VLA--11; the H\,{\sc ii} 
regions 1,\,2,\,3 
(Waller 1991); and the CO clouds 1,\,2,\,3 (Taylor et al.  1999) form a 
structure which resembles the 30 Doradus (R\,136) and N\,160\,$\&$N\,159 
region of the LMC (Cohen et al.  1988, Johansson et al 
1998, Bolatto et al.  2000), although  less massive.  Similar to the LMC, 
the domi\-nant CO clouds 1,\,2,\,3 (Taylor et al.  1999) are located at the 
Western end of NGC\,1569's stellar bar (Waller $\&$ Dracobly 1993) and extend 
from there in perpendicular direction to the major axis of the bar.  Star
formation has started at the Northern edge of the CO cloud, producing
the cluster No\,10 (Hunter et al.  2000) and intermediate-size clusters. 
The structure around cluster {\bf C} is difficult to assess; there are
several tentative sources M--1,\,b,\,c\,d, but for VLA
observations the region contains too much extended emission and
confusion to allow a clear distinction and identification of sources. 
The source M--1 is clearly a thermal source. 

Star formation is probably still progressing towards the South into the CO
cloud.  There are two non-thermal sources, M--2 and VLA--11, at the edge 
and $\sim$\,20\,--\,30\,pc outside  the CO cloud complex 1,\,2,\,3 
(Taylor et al.  1999), respectively.  If we interpret these sources as SNRs, 
or RSNe, we may conclude that at these positions half-way along and at the 
edge of the 
CO cloud, some star formation has taken place, or is currently taking place, 
although not (yet) very efficiently.  This holds in particular for the 
source M--2 since its surrounding contains the young, intermediate-size 
star cluster No\,5 of Hunter et al. (2000). 

There are two non-thermal sources to the North of cluster No\,10.  The source
VLA--16, at $\sim$\,25\,pc North, is probably a SNR, and possibly 
associated with the region of star cluster No\,10 ({\bf C}) or its
immediate surrounding (Table 4).  The same holds for the source M--3
which is associated with the cluster No\,8 of Hunter et al.  (2000; Table 4).  
In this area lies also the tentative source M--a.  

\subsection{The environment of the SSCs A and B}

The SSC--{\bf A} consists of two components (O'Connell et al. 1995, de Marchi 
et al.  1997, Hunter et al.  2000; see Table 4) and shows evidence of WR 
stars (Gonzalez--Delgado et al.  1997, Buckalew et al.  2000).  Under the 
assumption of being a single object, Ho $\&$ Filip\-penko (1996) derived a 
lower limit of the (dynamical) mass M({\bf A}) = 
(3.3\,$\pm$\,0.5)$\times$10$^{5}$M$_{\sun}$, based on the measured stellar 
velocity dispersion $\sigma$$_{*}$ = (15.7\,$\pm$\,1.5)\,km\,s$^{-1}$ and the 
half-light radius of the cluster r({\bf A}) = 1.9\,$\pm$\,0.2\,pc (0.18$''$).
Using the fact that {\bf A} is double, de Marchi et al. (1997) and Sternberg
(1998) obtain, for {\bf A1}, a mass between 2.8$\times$10$^{5}$M$_{\sun}$ and
1.1$\times$10$^{6}$\,M$_{\sun}$.  The \mbox{SSC(s)--{\bf A}} is located in a 
cavity, visible in 21\,cm-H\,{\sc i} and X-ray emission as a $\sim$\,200\,pc 
diameter hole (Israel $\&$ van Driel 1990, Heckman et al.  1995), The cavity 
is assumed to be blown out by SN explosions and stellar winds. The 
SSC--{\bf B} is located in some diffuse interstellar material.

In Fig.\,1, the circles around the clusters delineate areas of $\sim$\,50\,pc 
dia\-meter.  They represent, approximately, the distance a star escaping from 
a cluster with the velocity dispersion $\sigma$$_{*}$ $\approx$ 
20\,km\,s$^{-1}$ (Ho $\&$ Filippenko 1996) can traverse within 
$\sim$\,1\,--\,2\,Myr.  The encircled fields delineate approximately the areas
onto which a search for RSNe and SNRs, possibly associated with the SSCs, 
should concentrate.  The areas should not be significantly larger since the 
SSCs ({\bf A}) contain a large number of low mass stars and thus are 
likely to be gravitationally bound (Ho $\&$ Filippenko 1996, 
Sternberg 1998, Smith $\&$ Gallagher 2001) so that the probability that 
massive stars escape to larger distances is small, although the systems may 
not yet be fully relaxed.

\begin{table*}
\caption[]{Positions of star clusters referred to in this study
(Hunter et al. 2000; {\bf A1}, {\bf A2}: de Marchi et al. 1997).}
\begin{center}\small
\begin{tabular}{lcccrcc}
\hline
Cluster & RA\,(2000)  & Dec\,(2000) & $M_V^{a)}$ 
& \multicolumn{1}{c}{Radius$^{\ a)}$} & Non-thermal & Distance of RSN, SNR  \\
        & [h m s] & [$\degr$ $'$ $''$] & [mag]  
& \multicolumn{1}{c}{[$''$ $\leftrightarrow$ pc]} & Source & to Cluster [pc] \\
\hline
SSC--{\bf A} & 4 30 48.19 & 64 50 58.6 & --\,14.1 & 1.14\,--\,12\,\, & 
[VLA-8: SNR & $\sim$\,30] \\
SSC--{\bf A1}$^{\ b)}$ &       &      & --\,13.6 & $\sim$\,0.15\,--\,1.6 & & \\
SSC--{\bf A2}$^{\ b)}$ &    &            & --\,12.3 & $\sim$\,0.17\,--\,1.8 
& & \\ 
SSC--{\bf B} & 4 30 48.99 & 64 50 52.7 & --\,13.1 & 1.34\,--\,14\,\, & & \\
{\bf C}/No\,10 & 4 30 47.26 & 64 51 02.3 & --\,11.9 & 0.71\,--\,7.6 & VLA-16: 
SNR & $\sim$\,25 \\
No\,5       & 4 30 46.67 & 64 50 54.4 & --\,8.6 & 0.46\,--\,4.9 & M--2: 
RSN$^{\ c)}$ & 15 \\
No\,6       & 4 30 46.89 & 64 51 00.6 & --\,9.7 & 0.34\,--\,3.6 & M--1: therm. source &  \\
No\,7       & 4 30 46.96 & 64 50 59.4 & --\,9.2 & 0.34\,--\,3.6 & M--1: therm.
source & \\    
No\,8       & 4 30 47.04 & 64 51 06.6 & --\,8.6 & 0.23\,--\,2.4 & 
M--3: RSN$^{\ c)}$ & 2 \\
No\,18      & 4 30 48.07 & 64 50 57.3 & --\,7.8 & 0.18\,--\,1.9 & VLA-8: SNR 
& 10 \\ 
No\,45      & 4 30 54.53 & 64 50 43.2 & --\,6.9 & 0.50\,--\,6.0 & M--6: SNR 
& 35 \\ 
\hline
\multicolumn{7}{@{}l@{}}{%
a) for a distance of 2.2\,Mpc.}
\\
\multicolumn{7}{@{}l@{}}{%
b) components of SSC--{\bf A}, separated by 0.18$''$ [2.2\,pc]; {\bf A1} is 
located to the $\sim$\,South-East of {\bf A2}, Hunter et al. (2000).}
\\
\multicolumn{7}{@{}l@{}}{%
c) or a small SNR.}
\\
\end{tabular}
\end{center}
\end{table*}  

The observations do not reveal a RSN or SNR in the immediate surrounding
of the SSC {\bf A} and {\bf B}, except for the source VLA--8 assumed to be 
a SNR of $\sim$\,20\,pc diameter.

Besides the statistical argument brought forward to explain the absence 
of short-lived RSNe and RSNs in and near the SSCs, we believe that there 
exists also a valid kinematical argument for their absence.  When 
extrapo\-lating 
to SSCs Canto et al.'s (2000) calculation of the action of stellar winds of 
many massive stars in a cluster, combined with the action of several SN 
explosions, and when considering the influence of the cluster gravitational 
field on the propagation of the SN blast in a similar way as done for proto 
globular clusters (Shustov $\&$ Wiebe 2000), a violent and turbulent outflow 
of hot material is expected to occur which leaves little room for a quiescent 
development of SNRs.  Using the radius r({\bf A}) and mass M({\bf A}) of the 
SSC--{\bf A} mentioned above, the stellar mass concentration $\rho$ and the 
average distance $<$$\delta$$>$ between the cluster stars is

\begin{equation}
\rho \approx {M({\bf A}) \over {[4/3\pi {\rm r}({\bf A})^3]}} \approx
{3 \times 10^5 M_\odot \over {30 {\rm pc}^3}} \approx
1 \times 10^4 M_\odot {\rm pc}^{-3},
\end{equation}
and 

\begin{equation}
\langle \delta \rangle \propto (1/\rho)^{1/3} \propto 0.05 - 0.1 {\rm pc},
\end{equation}
if we assume that the average mass of the cluster stars is
$\sim$\,1\,--\,3\,M$_{\sun}$ (Sternberg 1998).  Stellar winds and material 
ejected in SN explosions extend to similar distances which makes a strong 
interaction of cluster-internal gas plausible.  An example of this process 
is SN\,1993J in M\,81 which has a shell diameter of $\sim$\,0.1\,pc some 
1300\,days after explosion and which expands with a velocity\footnote{in
a dense (molecular cloud) gas, like in M\,82, the 
expansion velocity can be significantly smaller (Chevalier $\&$ Fransson 
(2001).} of 
\mbox{$\sim$\,15\,000\,km\,s$^{-1}$} (Marcaide et al. 1997).  The individual 
stellar winds and SN ejected hot material blow out of the SSC as a common 
wind, which diffuses through the interstellar medium.  In the
immediate surroundings of the SSCs the outflowing stellar winds and SN
ejected material shock with the interstellar medium, forming shells and
holes.  The shock wave around SSC--{\bf A} has probably created the 
H\,{\sc i} hole and the shell seen in the observations of Israel $\&$ van 
Driel (1990) and Greve et al.  (1996).  In this picture it is not surprising 
that the non-thermal source (VLA--8), which may be an extended SNR, is
located $\sim$\,30\,pc outside the cluster core where the interstellar
matter is in less turbulent motion. This picture of the diffusion of SN 
ejected gas agrees with the fact that locally metal-enriched gas has not been
found (Kobulnicky $\&$ Skillman 1997). 

We do not find RSNe or SNRs in or near the many other star clusters (Hunter
et al. 2000) and WR sources (Buckalew et al. 2000), respectively.

\subsection{A final remark}
The phenomenon of SSCs and SNRs is much more spectacular in the starburst
galaxy M\,82 than in the irregular galaxy NGC\,1569; however, the heavy 
obscuration of M\,82 prevents us from obtaining a complete view of the 
relation between SSCs and SNRs. The post-starburst galaxy NGC\,1569 with 
locally recent star formation similar to that found in M\,82, allows on the 
other 
hand an unobscured view of many young stars in the body of the 
galaxy, of the large number of intermediate-size star clusters, and - as an 
exception to other unobscured closeby irregular galaxies - of two SSCs {\bf A} 
and {\bf B}. The VLA detects the non-thermal sources VLA--8 
and  VLA--16 at $\sim$\,25\,pc distance from the SSC {\bf A} and cluster 
{\bf C}, respectively. We did not detect these sources with MERLIN. We 
interpret these sources as low surface brightness 
SNRs. Because of their distance of $\sim$\,25\,pc from the clusters we hesitate
to attribute their origin to stars originally belonging to these clusters. 
The regions closer to the SSCs and closer to the intermediate-size clusters 
are devoid of RSNe and SNRs. Although a comparison between M\,82 and 
NGC\,1569 on statistical arguments seems to provide a plausible explanation 
for the absence of SNRs near and in the clusters, some caution in the use of 
this argument is appropriate in view of the locally large number 
of stars (10$^{5}$ to 10$^{6}$) very recently formed in the SSCs and the 
intermediate-size clusters. The absence of SNRs in and very close to the 
clusters, in both M\,82 and NGC\,1569, may -- at least partially -- 
be due to the hostile environment. Unfortunately, within $\sim$\,5\,Mpc 
distance there are no other unobscured galaxies containing many SSCs, 
allowing a similar investigation with MERLIN.

\begin{acknowledgements} We thank the MERLIN staff, Jodrell Bank, for the
observations, the help in data reduction, and the pleasant hospitality.
We thank the referee for putting the astrophysical question into the correct
context of star formation and supernova explosions, and for eliminating 
contradictions.
\end{acknowledgements}

\end{document}